\documentclass[twocolumn,showpacs,amsmath,amsfonts,amssymb,superscriptaddress,nofootinbib,preprintnumbers]{revtex4-1}

\pdfoutput=1
\usepackage[usenames,dvipsnames]{color}
\usepackage[colorlinks=true, linkcolor=BrickRed, citecolor=Blue, urlcolor=Blue, filecolor=Blue]{hyperref}
\usepackage{longtable,epsfig,graphicx,verbatim,xspace,multirow,ulem,hhline}

\begin{document}

\preprint{CFTP/16-05} 	
\preprint{IFIC/16-07}

\title{Isotropic extragalactic flux from dark matter annihilations: \\
	lessons from interacting dark matter scenarios}

\author{\'Angeles Molin\'e}
\affiliation{CFTP, Instituto Superior Tecnico, Universidade Tecnica de Lisboa, Av. Rovisco Pais 1, 1049-001 Lisboa, Portugal}
\affiliation{Instituto de F\'{\i}sica Corpuscular (IFIC),  CSIC-Universitat de Val\`encia,  Apartado de Correos 22085, E-46071 Valencia, Spain}
\author{Jascha A. Schewtschenko}
\affiliation{Institute for Particle Physics Phenomenology (IPPP), Durham University, Durham DH1 3LE, UK}
\affiliation{Institute for Computational Cosmology (ICC), Durham University, Durham DH1 3LE, UK}
\author{Sergio Palomares-Ruiz}
\affiliation{Instituto de F\'{\i}sica Corpuscular (IFIC),  CSIC-Universitat de Val\`encia,  Apartado de Correos 22085, E-46071 Valencia, Spain}
\author{C\'eline B\oe hm}
\affiliation{Institute for Particle Physics Phenomenology (IPPP), Durham University, Durham DH1 3LE, UK}
\affiliation{LAPTH, U. de Savoie, CNRS, BP 110, 74941 Annecy-Le-Vieux, France}
\author{Carlton M. Baugh}
\affiliation{Institute for Computational Cosmology (ICC), Durham University, Durham DH1 3LE, UK}

%%%%%%%%%%%%%%%%%%%%%%%%%%%%%%%%%%%%%%%%%%%%%%%%%%%%%
%%%%%%%%%%%%%%%%%%%%%%%%%%%%%%%%%%%%%%%%%%%%%%%%%%%%%

\begin{abstract}
The extragalactic $\gamma-$ray and neutrino emission may have a contribution from dark matter (DM) annihilations. In the case of discrepancies between observations and standard predictions, one could infer the DM pair annihilation cross section into cosmic rays by studying the shape of the energy spectrum.  So far all analyses of the extragalactic DM signal have assumed the standard cosmological model ($\Lambda$CDM) as the underlying theory. However, there are alternative DM scenarios where the number of low-mass objects is significantly suppressed. Therefore the characteristics of the $\gamma-$ray and neutrino emission in these models may differ from $\Lambda$CDM as a result. Here we show that the extragalactic isotropic signal in these alternative models has a similar energy dependence to that in $\Lambda$CDM, but the overall normalisation is reduced. The similarities between the energy spectra combined with the flux suppression could lead one to misinterpret possible evidence for models beyond $\Lambda$CDM as being due to CDM particles annihilating with a much weaker cross section than expected.  
\end{abstract}

\date{\today}

\maketitle

%%%%%%%%%%%%%%%%%%%%%%%%%%%%%%%%%%%%%%%%%%%%%%%%%%%%%
%%%%%%%%%%%%%%%%%%%%%%%%%%%%%%%%%%%%%%%%%%%%%%%%%%%%%
\section{Introduction}
\label{sec:intro}
%%%%%%%%%%%%%%%%%%%%%%%%%%%%%%%%%%%%%%%%%%%%%%%%%%%%%
%%%%%%%%%%%%%%%%%%%%%%%%%%%%%%%%%%%%%%%%%%%%%%%%%%%%%

Finding signatures of  dark matter (DM) annihilation or decay in the sky is key to establishing its microscopic nature.  One promising avenue is the detection of an anomalous population of cosmic rays and electromagnetic emission in DM haloes. Haloes of well identified dwarf galaxies and clusters of galaxies have been targeted in $\gamma-$ray~\cite{Ackermann:2010rg, Ackermann:2012nb, Ackermann:2013yva, Ackermann:2015zua, Buckley:2015doa, Ahnen:2016qkx, Albert:2007xg, Aliu:2008ny, MAGIC:2009, Aleksic:2011jx, Aleksic:2013xea, Abramowski:2012au, Pfrommer:2012mm, Aharonian:2007km, Aharonian:2008dm, Abramowski:2010aa, Abramowski:2014tra, Acciari:2010ab, Aliu:2012ga} and neutrino~\cite{Sandick:2009bi, Yuan:2010gn, Dasgupta:2012bd, Murase:2012rd, Aartsen:2013dxa} searches. No anomalous signal has been found yet. The isotropic diffuse signal that originates from all mass DM haloes, including those outside our cosmic neighbourhood, is now also under close scrutiny~\cite{Beacom:2006tt, Murase:2012xs, Moline:2014xua, Ackermann:2015tah}.

While the intensity of the flux expected from individual haloes for a given DM mass depends mostly on the DM density profile in the targeted region and is fairly straightforward to compute, estimates of the extragalactic emission are much more uncertain and challenging. Not only do the predictions require a good knowledge of the internal structure of haloes, but they  also require the determination of their abundance in the past Universe, and at all possible scales.

The halo abundance can be predicted using N-body simulations or can be inferred from the linear perturbation theory matter power spectrum using semi-analytical techniques\footnote{Analytical calculations are calibrated to the output of N-body simulations which provide a way to assess the validity of the results.}. Different analytical models have been developed~\cite{Press:1973iz, Colafrancesco:1989px, Peacock:1990zz, Appel:1990, Bond:1990iw, Bower:1991kf, Lacey:1993iv, Kauffmann:1993jn, Lacey:1994su, Monaco:1994ed, Manrique:1995im, Manrique:1995hf, Lee:1997cj, Manrique:1997eg, Sheth:1999mn, Sheth:1999su, Sheth:2001dp, Maggiore:2009rv, Maggiore:2009rw, Maggiore:2009rx, Paranjape:2012ks, Juan:2014xra} and a number of parameterisations have been  suggested to fit the simulations~\cite{Jenkins:2000bv, Reed:2003sq, Warren:2005ey, Reed:2006rw, Lukic:2007fc, Cohn:2007xu, Tinker:2008ff, Crocce:2009mg, Courtin:2010gx, Bhattacharya:2010wy, Angulo:2012ep, Watson:2012mt, Rodriguez-Puebla:2016ofw}. However, the resolution of the current simulations is not sufficient to establish the halo abundance at very low masses (corresponding to small smoothing scales) and high redshifts, a limitation which could hinder flux estimates. A solution to compute the extragalactic flux stemming from DM annihilations is to extrapolate the halo abundance found at high masses using numerical or semi-analytical techniques, down to the smallest mass scales.

However, this approach can generate large uncertainties since the presence of low-mass haloes in galactic haloes (and in the Universe in general) is under debate. This is referred to as the ``missing satellite" problem in the standard model of cosmology, $\Lambda$CDM, which is based on a cosmological constant and cold dark matter (CDM) and seems to over predict the existence of such objects~\cite{Klypin:1999uc, Moore:1999nt}. While the current discrepancy with observations could be due to the baryonic processes affecting galaxy formation such as feedback~\cite{Sawala:2015cdf}, a number of alternative DM models has been proposed to solve this issue as well as the ``too big to fail", ``cusp vs. core" and ``satellite alignment" problems~\cite{Goetz:2002vm, Vogelsberger:2012ku}.  A simpler explanation however (if feedback is not the right solution) could be that these anomalies are a manifestation of the impact of DM interactions with other particles on the form of the primordial matter fluctuations~\cite{Boehm:2000gq, Boehm:2001hm, Boehm:2004th}. 

In standard $\Lambda$CDM scenarios, where heavy and weakly interacting massive particles (WIMPs) are the prime DM candidates, such interactions are known to have a negligible effect on structure formation. In this case, the impact of a late kinetic decoupling time on the free-streaming scale leads to a cut-off scale in the linear matter power spectrum between $M_{\rm min}=10^{-12}-10^{-4}M_\odot$~\cite{Schmid:1998mx, Zybin:1999ic, Boehm:2000gq, Chen:2001jz, Hofmann:2001aa, Berezinsky:2003vn, Boehm:2003aa, Boehm:2004th, Green:2005fa, Loeb:2005pm, Profumo:2006bv, Bertschinger:2006nq, Bringmann:2006mu, Bringmann:2009vf, vandenAarssen:2012ag, Cornell:2012tb, Gondolo:2012vh, Cornell:2013rza, Shoemaker:2013tda, Diamanti:2015kma}. So all cosmologically relevant structures, including the smallest ones, are expected to contribute to the diffuse extragalactic emission in CDM models.

The more general approach taken in Refs.~\cite{Boehm:2000gq, Boehm:2004th} however, shows that there are many more possible DM models, some of which predict a different pattern of structure formation due to a large collisional damping effect, even though they are WIMP candidates~\cite{Boehm:2003aa, Boehm:2003hm}. For example, if DM particles interact with photons or neutrinos with a cross section $\sigma_{\rm el} \simeq 10^{-33} \, \left( m_{\rm DM}/{\rm{GeV}}\right) \, {\rm cm}^2$, one expects a cut-off in the linear matter power spectrum around a smooth scale of $\sim$100~kpc~\cite{Boehm:2001hm, Sigurdson:2004zp, Mangano:2006mp, Serra:2009uu, Wilkinson:2013kia, Wilkinson:2014ksa, Cyr-Racine:2013fsa}, and therefore a suppression of the number of haloes smaller than those which host dwarf galaxies. The results are similar if the interactions occur with baryons or with the DM itself, but the cross section has to be about ten orders of magnitude larger than that with photons or neutrinos to produce similar results~\cite{Boehm:2004th, Chen:2002yh, Dvorkin:2013cea}. 

DM interactions lead to a different linear perturbation theory matter spectrum than the case of warm dark matter (WDM). The matter power spectrum, $P(k)$, of interacting dark matter (IDM) features damped oscillations, referred to as {\it dark oscillations}, below the collisional damping cut-off scale, while the equivalent power spectrum for a thermal WDM relic can be approximated by a steep exponential cut-off (see, e.g., Ref.~\cite{Bode:2000gq}). The effect of an oscillating $P(k)$ on large scale structures and on the Milky-Way-size haloes was simulated in Refs.~\cite{Boehm:2014vja, Schewtschenko:2014fca, Vogelsberger:2015gpr, Cyr-Racine:2015ihg}.  The results show that such a suppression in $P(k)$ can actually explain the ``missing satellite" and ``too big to fail" problems~\cite{Schewtschenko:2015rno}.  However, most importantly for our purposes, these simulations provide us with the halo abundance down to the scale of dwarf galaxies and at large redshifts, which allows us to explore the impact of such a damped linear matter power spectrum on the estimate of the isotropic cosmological fluxes of $\gamma-$rays and neutrinos from DM annihilations. 

In Section~\ref{sec:xgal} we summarize the main ingredients entering the calculation of the isotropic cosmological fluxes of the products of DM annihilations. In Section~\ref{sec:alternative} we describe IDM scenarios and their cosmological signatures. The description of the simulations used in this work is given in Section~\ref{sec:simulations} and the results presented in Section~\ref{sec:results}. Finally, we draw our conclusions in Section~\ref{sec:conclusions}.

%%%%%%%%%%%%%%%%%%%%%%%%%%%%%%%%%%%%%%%%%%%%%%%%%%%%%
%%%%%%%%%%%%%%%%%%%%%%%%%%%%%%%%%%%%%%%%%%%%%%%%%%%%%
\section{Extragalactic emission from DM annihilations}
\label{sec:xgal}
%%%%%%%%%%%%%%%%%%%%%%%%%%%%%%%%%%%%%%%%%%%%%%%%%%%%%
%%%%%%%%%%%%%%%%%%%%%%%%%%%%%%%%%%%%%%%%%%%%%%%%%%%%%

Here we summarize some well-known general expressions and definitions for the calculation of the isotropic extragalactic $\gamma-$ray and neutrino flux from DM annihilations, which apply to the different scenarios we consider.

 \subsection{Extragalactic $\gamma-$ray flux}
  
The contribution to the diffuse $\gamma-$ray emission arising from the annihilation of DM particles in a single halo of mass $M$ at a given redshift $z$ is given by\footnote{Note that this expression is valid when DM particles and antiparticles are identical. Otherwise, an extra factor of $1/2$ must be added.}~\cite{Ullio:2002pj, Taylor:2002zd}
\begin{eqnarray}
\frac{dn_{\gamma}(E_0)}{dE_{0}}\Big|_{M,z} & = & \frac{1}{4\pi \chi^2(z)}
\frac{\langle\sigma \upsilon\rangle}{2 \, m^2_{\rm DM}} \, \left< \rho^2 \right>\Big|_{M,z}  \times \nonumber \\
& & \, e^{-\tau(E_{0},z)} \, \sum_{i} {\rm Br}_{i}
\frac{dN_{i}(E)}{dE} ~,    
\label{eq:dndE}
\end{eqnarray}
where $\chi(z)$ is the comoving distance of the halo, $\langle\sigma \upsilon\rangle$ is the averaged DM annihilation cross section times relative velocity of the pair, $m_{\rm DM}$ is the DM mass, and $\left< \rho^2 \right>\Big|_{M,z} = \int 4\pi r^2 \rho^2(r,M,z) \, dr$ is proportional to the enhancement for a single halo of density $\rho$ with respect to a smooth distribution. The optical depth of attenuation of $\gamma-$rays in the extragalactic background light is given by $\tau(E_{0},z)$, and we adopt the model of Ref.~\cite{Dominguez:2010bv}. $dN_{i}/dE$ is the differential $\gamma-$ray energy spectrum per annihilation into the channel $i$, with branching ratio ${\rm Br}_{i}$. In Eq.(\ref{eq:dndE}), $E_0$ is the energy at the Earth and $E=E_0 (1+z)$ is the energy at the source. To calculate the spectrum at the source, we make use of the results presented in Ref.~\cite{Cirelli:2010xx}, which were computed using PYTHIA 8.1~\cite{Sjostrand:2007gs} and include electroweak corrections~\cite{Ciafaloni:2010ti}. 

Taking into account the number density of haloes per unit mass at different redshifts, the contribution of all haloes at all  cosmological distances is given by
\begin{eqnarray}
\frac{d \phi_{\gamma} (E_0)}{dE_{0}} & = & \frac{\langle\sigma
	\upsilon\rangle}{2} \frac{\rho^{2}_{\rm m,0}}{m_{\rm DM}^{2}} \, \int \frac{dz}{H(z)} \, \xi^{2}(z) \times \nonumber \\
& & e^{-\tau(E_{0},z)} \, \sum_{i} {\rm Br}_{i} \frac{dN_{\gamma, i}(E_{0}(1+z))}{dE} ~,    
\label{eq:fluxgamma}
\end{eqnarray}
where $H(z)= H_0 \, \sqrt{\Omega_{\rm m,0}(1+z)^{3} +  \Omega_{\Lambda}} \equiv H_0 \, h(z)$ is the Hubble parameter as a function of redshift, $\rho_{\rm m, 0}$ is the comoving matter density today and $\xi^2(z)$ is the enhancement factor at a given redshift, which is defined in Section~\ref{sec:xi}.

\subsection{Extragalactic neutrino flux}

In the case of neutrinos, there are two main differences in the calculation of the isotropic extragalactic flux with respect to the case of $\gamma-$rays. Firstly, neutrinos interact only via weak interactions and hence, they traverse cosmic distances without any significant absorption. Secondly, neutrinos experience flavour oscillations. Whereas they are produced in some combination of flavour eigenstates, propagation over cosmic distances results in averaged flavour oscillations. Therefore, the integrated neutrino flux from DM annihilations in all haloes of all masses at all redshifts is given by
\begin{eqnarray}
\frac{d\phi_{\nu_\alpha} (E_0)}{dE_0} & = & \frac{\langle\sigma \upsilon\rangle}{2} \frac{\rho^{2}_{\rm m,0}}{m_{\rm DM}^{2}} \, \int \frac{dz}{H(z)} \, \xi^{2}(z) \times  \\
& & \sum_{\beta,i} |U_{\alpha,i}|^2 \, |U_{\beta,i}|^2 \, \sum_{i} {\rm Br}_{i} \frac{dN_{\nu_\beta, i}(E_{0}(1+z))}{dE}   ~,  \nonumber  
\label{eq:fluxnu}
\end{eqnarray}
where $dN_{\nu_\beta, i}/dE$ is the flux of neutrinos of flavour $\beta$ at the source  after DM annihilations into channel $i$, and $U$ is the leptonic mixing matrix. For the analysis, we use the latest $\nu$\textit{fit} results~\cite{GonzalezGarcia:2012sz} (see also Refs.~\cite{Tortola:2012te, Fogli:2012ua}).

\subsection{Enhancement factor $\xi^2(z)$}
\label{sec:xi}

The enhancement factor $\xi^2(z)$ in Eqs.~(\ref{eq:fluxgamma}) and~(\ref{eq:fluxnu}) is given by~\cite{Taylor:2002zd}  
\begin{equation}
\xi^2(z)= \frac{\Delta(z) \, \rho_{\rm c}(z)}{\rho_{\rm m,0}} \,
\int_{M_{\rm min}} dM \frac{M}{\rho_{{\rm m},0}}\frac{dn(M,z)}{dM} \, \xi^2_{\rm M}(M,z) ~, 
\label{eq:xi2}
\end{equation}
where $\Delta$ is the overdensity parameter used to define spherical haloes ($M \equiv M_\Delta = 4\pi/3 \, \Delta \rho_c(z) \, R_\Delta^3$) which we set to $\Delta=200$ (see Section~\ref{sec:simulations}), $\rho_c(z)$ is the critical density of the Universe at redshift $z$, $dn(M,z)/dM$ is the halo mass function for haloes of mass $M$ at redshift $z$ and $\xi^2_{\rm M}(M,z)$ is the enhancement of the signal of a single halo, and is defined as~\cite{Moline:2014xua}
\begin{equation}
\xi^2_{\rm M}(M,z) = \frac{M}{\Delta \, \rho_c(z)} \, \frac{\left< \rho^2 \right>}{\left< \rho \right>^2} ~.
\end{equation}
$\xi^2_{\rm M}$ represents the enhancement in flux compared to that produced by a smooth DM distribution with density $\Delta(z) \, \rho_{\rm c}(z)$, in a volume $4\pi R_\Delta^3/3$. For a Navarro-Frenk-White (NFW) profile~\cite{Navarro:1995iw,Navarro:1996gj}, $\xi^2_{\rm M}$ has an analytical form in terms of the concentration parameter, $c=R_\Delta/r_s$ ($r_s$ is the scale radius), given by~\cite{Moline:2014xua}
\begin{equation}
\xi^2_{\rm M}(c(M),z)\Big|_{\rm NFW} = \frac{1}{9} \,
\frac{{c}^{3} \left(1-(1+c)^{-3}\right)}{\left[ \ln(1+c)-c (1+c)^{-1} \right]^{2}} ~.    
\label{eq:xiM2NFW}
\end{equation}
%

%%%%%%%%%%%%%%%%%%%%%%%%%%%%%%%%%%%%%%%%%%%%%%%%%%%%%
%%%%%%%%%%%%%%%%%%%%%%%%%%%%%%%%%%%%%%%%%%%%%%%%%%%%%
\section{Alternative DM models and power spectra}
\label{sec:alternative}
%%%%%%%%%%%%%%%%%%%%%%%%%%%%%%%%%%%%%%%%%%%%%%%%%%%%%
%%%%%%%%%%%%%%%%%%%%%%%%%%%%%%%%%%%%%%%%%%%%%%%%%%%%%

The standard model of cosmology assumes that DM is well described by a collisionless fluid. In practice, this means that one assumes that the DM interactions are far too weak to have any effect on structure formation. However, in most models, DM must have non negligible interactions (with standard model particles or with particles in the dark sector) to explain the observed DM abundance. It is therefore important to determine the impact of such interactions on the formation of objects in the Universe.  

DM interactions have two main effects on the primordial DM fluctuations: first, they induce a collisional damping (a generalisation of the Silk damping effect); secondly, they delay the free-streaming of the DM particles~\cite{Boehm:2000gq, Boehm:2004th}.  Fluctuations are therefore first erased by the DM scattering with standard model (or dark sector, including the DM itself) particles and the scales that have not been erased by the collisions may eventually be erased by the DM free-streaming. 

The strength of the effect is governed by the ratio of the elastic cross section of DM with a given species to the DM mass~\cite{Boehm:2001hm, Sigurdson:2004zp, Mangano:2006mp, Serra:2009uu, Wilkinson:2013kia, Cyr-Racine:2013fsa, Wilkinson:2014ksa}. The larger the ratio, the larger the effect of  collisional damping. The characteristics of the species the DM interacts with (energy density and velocity) play a very important role. For a fixed value of the cross section-to-mass ratio, the collisional damping effect is the largest if DM interacts with photons or neutrinos rather than with itself or with baryons, since the energy densities and velocities of photons and neutrinos are much larger than those of the baryons or the DM particles.

The collisional damping of primordial DM fluctuations induces a damped oscillating linear matter power spectrum with a cut-off scale that is essentially given by the collisional damping scale. This gives rise to an exponential cut-off in the linear $P(k)$, like for WDM. However, the competition between pressure due to collisions and gravity generates, at smaller scales, an oscillatory pattern in the matter power spectrum, like that seen in the case of Silk damping~\cite{Silk:1967kq}.

Numerical simulations of the impact of these interactions on structure formation have been performed in Refs.~\cite{Schewtschenko:2014fca, Boehm:2014vja}. It was shown that even a weak DM elastic scattering cross section with neutrinos or photons could have a huge impact on Milky Way-like haloes and smaller structures, if DM is lighter than a few GeV. Typically, a cross section larger than $\sigma_{\rm el} \gtrsim 10^{-33} \ \left(m_{\rm{DM}}/\rm{GeV}\right) \  \rm{cm^2}$ would erase all dwarf galaxies in a Milky Way-like halo. This constraint is expected to become more stringent with the next generation of large-scale-surveys~\cite{Escudero:2015yka}, like DESI~\cite{Levi:2013gra} and LSST~\cite{Abell:2009aa}, and the inclusion of baryonic physics in the simulations, thus demonstrating the importance of taking into account even small interactions, which are relevant for the relic density calculations. 

In the total absence of DM interactions (the pure collisionless case) and assuming light DM particles, the DM free-streaming scale is $l_{\rm fs}\propto m_{\rm DM}^{-4/3}$~\cite{Bode:2000gq, Boehm:2000gq, Boehm:2004th}. To make sure that small objects such as dwarf galaxies can form (i.e., $l_{\rm fs} \lesssim 100$ kpc), thermally produced DM particles must be heavier than a few keV; a conclusion which is also in agreement with Lyman-alpha forest analyses~\cite{Viel:2013apy}.  Scenarios featuring keV particles are examples of collisionless WDM scenarios. Collisionless candidates heavier than a few keV, and produced thermally\footnote{However, see, for instance, Ref.~\cite{Gelmini:2009xd} for warmer-than-thermal candidates.}, are well described by CDM (with a different expression for the free-streaming scale~\cite{Boehm:2000gq}). In the presence of interactions (the case of IDM) the expression for the free-streaming scale is different and depends on both the DM mass and the interaction cross section. Besides, even for small cross sections, the collisional damping scale could still dominate over the free-streaming scale.

In general, IDM models are very similar to WDM, even though the suppression of power in the linear matter power spectrum is not as drastic as with WDM due to the oscillations. Therefore, the isotropic extragalactic flux produced by the products of DM annihilations is expected to be similarly suppressed in both cases. Thus, the question we want to address is whether these alternative matter power spectra could lead to very different $\gamma-$ray or neutrino fluxes than those predicted in the standard CDM scenario. 

To compute the extragalactic emission in IDM scenarios, we consider their halo mass function and concentration-mass relation as a function of redshift and assume the canonical value for the DM annihilation cross section into standard model particles. For IDM, we assume that the DM annihilation and scattering cross sections are not connected so that the matter power spectrum serves as a template of alternative $\Lambda$CDM scenarios. From the simulations, we use the information on the structures that are present in the redshift interval $z = [0, \, 4]$. The larger the redshift, the more suppressed the small scale structures~\cite{Boehm:2003xr}. Hence, in these alternative DM models, the main contribution to the signal is expected to come from very late times, i.e., from even lower redshifts than in the standard scenario. In principle, this could induce differences in the shape of the $\gamma-$ray and neutrino spectra.

%%%%%%%%%%%%%%%%%%%%%%%%%%%%%%%%%%%%%%%%%%%%%%%%%%%%%
%%%%%%%%%%%%%%%%%%%%%%%%%%%%%%%%%%%%%%%%%%%%%%%%%%%%% 
\section{Simulations} 
\label{sec:simulations}
%%%%%%%%%%%%%%%%%%%%%%%%%%%%%%%%%%%%%%%%%%%%%%%%%%%%%
%%%%%%%%%%%%%%%%%%%%%%%%%%%%%%%%%%%%%%%%%%%%%%%%%%%%%

We use the output of the DM  simulations performed in Ref.~\cite{Schewtschenko:2014fca} to account for DM-radiation interactions,  using the parallel $N$-body TreePM code \texttt{GADGET-3}~\cite{Springel:2005mi}.  The simulations were performed from $z=49$ to $z=0$, and assumed that interactions were negligible at late times ($z<49$). The matter power spectrum at $z=49$ was obtained from a modified version of the Boltzmann code \texttt{CLASS}~\cite{Wilkinson:2013kia}, using the best fitting values of the cosmological parameters obtained by the {\it Planck} collaboration,  using the ``{\it Planck} + WP'' dataset~\cite{Ade:2013zuv} and assuming a $\Lambda$CDM cosmology. In principle, a consistent treatment of an interacting DM model would require one to use the best fitting cosmology for a given scattering cross section. However, the parameters for $\Lambda$CDM lie well within the one standard deviation of such best fits. Therefore one can use the same values of the cosmological parameters as given by the {\it Planck} collaboration to perform the IDM simulations. The initial conditions were created with an adapted version of a second-order Lagrangian perturbation theory code~\cite{Crocce:2006ve}.

To provide a suitable dynamical range, simulations in both a large volume ($100~{h}^{-1}~{\rm Mpc}$) and a small volume ($30~{h}^{-1}~{\rm Mpc}$) were run with $1024^3$ DM particles. The gravitational softening was set to 5$\%$ of the mean particle separation to avoid artificially enhanced two-body relaxation processes between the particles. 

For the identification of the dark matter haloes in the simulations, we used the \texttt{AMIGA} halo finder~\cite{Knollmann:2009pb}, which identifies collapsed structures as spherically overdense regions of radius $R_{200}$ with a mean density given by
\begin{equation}
\frac{3 \, M_{200}}{4 \, \pi \, R^3_{200}} = 200 \, \rho_c(z) ~,
\label{eq:sim:halodef}
\end{equation}
where $M_{200}$ is the halo mass\footnote{This definition of a halo differs slightly from the one used in Ref.~\cite{Schewtschenko:2014fca}, so the measured halo properties may show some differences from the results presented there.}.

We then averaged the density in shells around the centre-of-mass for all haloes in a given mass bin, as done in Ref.~\cite{Schewtschenko:2014fca} to determine the density profiles. The resulting density distribution is found to be in good agreement with the NFW parameterisation, which is completely characterised by the concentration parameter. The latter is in turn  determined by using the approach described in Ref.~\cite{Prada:2011jf}.

Combining the results of the two simulation volumes with their respective resolutions and, also binning the DM haloes by mass, we were  able to determine a median concentration parameter $c_{200}(M_{200},z)$ for the following redshift and mass intervals

\begin{itemize}
	\item for $z\le1$:  $9.6<\log_{10} M_{200} \, [h^{-1} M_{\odot}]<14.8$ ~, 
	\item for $z=2$: $9.6<\log_{10} M_{200} \, [h^{-1} M_{\odot}]<14$ ~, 
	\item for $z=3$:  $9.6<\log_{10} M_{200} \, [h^{-1} M_{\odot}]<13.6$ ~,    
	\item for $z=4$:  $9.6<\log_{10} M_{200} \, [h^{-1} M_{\odot}]<13.2$ ~,             
\end{itemize}
for the models with DM-radiation interactions. The declining upper limit with increasing redshift simply reflects the fact that no structure above that mass range had formed in the simulation volume. The  lower limit is chosen so as to avoid contamination by spurious haloes~\cite{Schewtschenko:2014fca}, which would otherwise distort the concentration-mass relation. Such haloes are a feature in simulations of models with damped input spectra (see, e.g., Ref.~\cite{Wang:2007he}).

In what follows, we show results for the case of elastic scatterings of DM particles off photons with a cross section $\sigma_{\rm el} = 2.0 \times 10^{-9} \, (m_{\rm DM}/\rm{GeV}) \times \sigma_{\rm T}$ ($\sigma_{\rm T}$ is the Thomson cross section).

%%%%%%%%%%%%%%%%%%%%%%%%%%%%%%%%%%%%%%%%%%%%%%%%%%%%%
%%%%%%%%%%%%%%%%%%%%%%%%%%%%%%%%%%%%%%%%%%%%%%%%%%%%%
\section{Results}
\label{sec:results}
%%%%%%%%%%%%%%%%%%%%%%%%%%%%%%%%%%%%%%%%%%%%%%%%%%%%%
%%%%%%%%%%%%%%%%%%%%%%%%%%%%%%%%%%%%%%%%%%%%%%%%%%%%%

We now describe the extragalactic fluxes in the IDM scenario we consider and discuss the main differences with respect to the standard CDM case.

\subsection{The halo mass function}

The number density of haloes per unit mass as a function of mass and redshift is quantified by the comoving halo mass function~\cite{Jenkins:2000bv},    
\begin{equation}
\frac{dn(M,z)}{dM} = \frac{\rho_{\rm m,0}}{M^{2}} \,
\frac{d\,\mbox{ln}\sigma^{-1}}{d\,\mbox{ln} M} \, f(\sigma(M,z)) ~, 
\label{eq:dndm}
\end{equation}
where the function $f (\sigma)$ represents the fraction of mass in collapsed haloes per unit interval in $\ln \sigma^{-1}$ and, if all the mass is inside haloes, it satisfies $\int f(\sigma)  d\ln{\sigma^{-1}} = 1$. The variance of the linear density field, $\sigma(M,z)$, is given by
\begin{equation}
\sigma^{2}(M,z) = \left(\frac{D(z)}{D(0)}\right)^{2} \int \frac{dk}{k}\frac{k^{3} \, P(k)}{2\pi^{2}}
|\hat{W}(kR)|^{2} ~,
\label{eq:sigma} 
\end{equation} 
where $D(z)$ is the growth factor, $P(k)$ is the matter power spectrum which is smoothed on a mass-dependent scale, $R(M)^3 = 3 \, M/4\pi \rho_{\rm m,0}$, with a spherical top-hat window, $W(r,R)$, whose Fourier transform is $\hat{W}(kR)$. For the alternative IDM scenarios  considered here, where there is a strong suppression in power at small scales, this window function does not reproduce correctly the halo mass function~\cite{Schneider:2011yu, Schewtschenko:2014fca}. A mass-dependent correction to the Sheth-Tormen halo mass function~\cite{Sheth:1999mn, Sheth:1999su, Sheth:2001dp} could, in principle, solve this discrepancy~\cite{Schneider:2011yu},
\begin{equation}
\frac{dn(M,z)}{dM}= \left(1+\frac{M_{\rm hm}}{\beta M}\right)^{-\alpha}\frac{dn_{\rm IDM, ST}(M,z)}{dM} ~,
\label{eq:HMFWDM}
\end{equation}
where $M_{\rm hm} = 4.3 \times 10^9 \, h^{-1} \, M_\odot$ is the half mode mass, the scale at which the transfer function is half that for CDM, and $\alpha = 0.6$ and $\beta=0.5$ are free parameters\footnote{Notice that in Ref.~\cite{Schewtschenko:2014fca} there is a typo in the value quoted for $\beta$.} which were found after fitting the results of the simulations for $M_{200} > 10^9 \, h^{-1} \, M_\odot$~\cite{Schewtschenko:2014fca}, for the scattering cross section used here. Note that the subindex $_{\rm IDM,ST}$ in Eq.~(\ref{eq:HMFWDM}) indicates the use of the IDM power spectrum and the Sheth-Tormen parameterisation. Moreover, we have used a different criterion to identify halos from that used in Ref.~\cite{Schewtschenko:2014fca}, so we have also used a different normalisation, $A=0.285$, from the one usually quoted for the Sheth-Tormen halo mass function for CDM, $A=0.322$.

However, as seen in Ref.~\cite{Schewtschenko:2014fca}, the aforementioned correction term, which works for WDM~\cite{Schneider:2011yu}, does not reproduce the IDM results for masses below $M \simeq 10^9 \, h^{-1} \, M_\odot$. An alternative correction term is given by
\begin{equation}
\frac{dn(M,z)}{dM}= \left(1+\frac{M_{\rm hm}}{\beta M}\right)^{\alpha}\left(1+\frac{M_{\rm hm}}{\gamma M}\right)^{\delta}\frac{dn_{\rm CDM, ST}(M,z)}{dM} ~,
\label{eq:HMFIDM}
\end{equation}
where $\alpha=-1.0$, $\beta=0.33$, $\gamma=1$ and $\delta=0.6$ are found by fitting  the results of IDM simulations. Now, the subindex $_{\rm CDM,ST}$ refers to the usual Sheth-Tormen parameterisation for the CDM power spectrum, but with  $A=0.285$.

\begin{figure}[t]
	\centering	
	\includegraphics[width=0.5\textwidth]{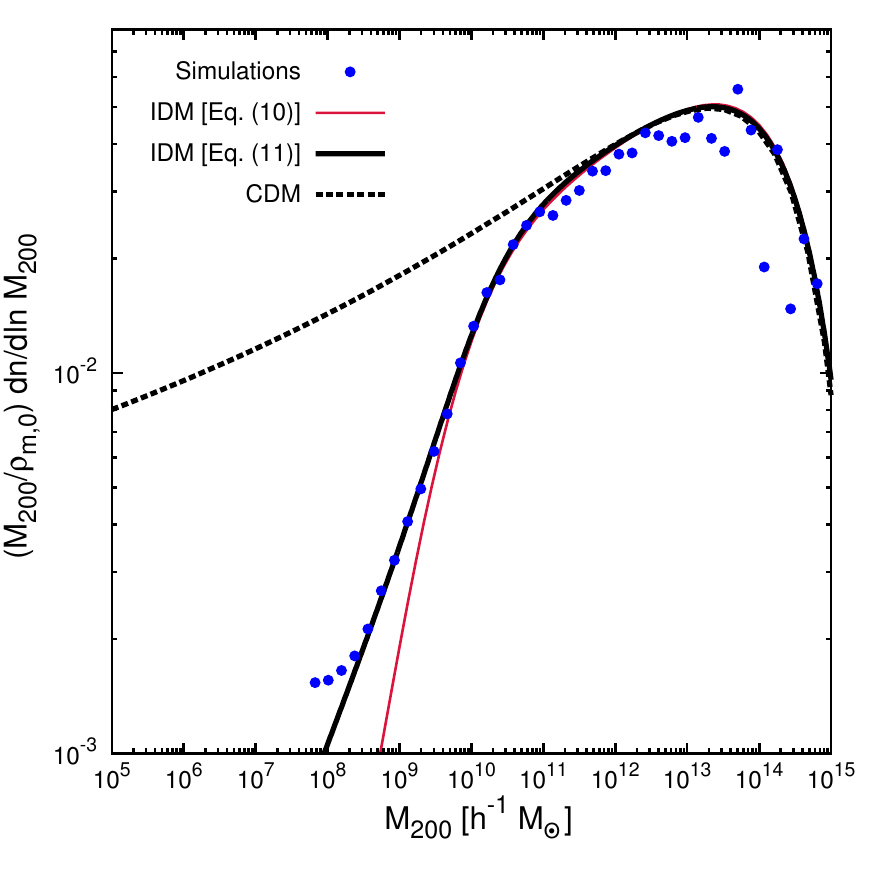}
	\caption{Halo mass function at $z=0$ for IDM with an elastic DM-$\gamma$ (similarly for DM-$\nu$) scattering cross section $\sigma_{\rm el} = 2.0 \times 10^{-9} \, (m_{\rm DM}/\rm{GeV}) \times \sigma_{\rm T}$ (blue dots). The fitting functions in Eq.~(\ref{eq:HMFWDM}) (solid red line) and Eq.~(\ref{eq:HMFIDM}) (solid black line), and the Sheth-Tormen parameterisation for the standard CDM scenario~\cite{Sheth:1999mn, Sheth:1999su, Sheth:2001dp} (dashed black line) with normalisation $A=0.285$, are also shown.}
	\label{fig:hmf} 
\end{figure}

In Fig.~\ref{fig:hmf} we show the halo mass function at $z = 0$ and compare it with the standard result from CDM, for which we have used the Sheth-Tormen parameterisation~\cite{Sheth:1999mn, Sheth:1999su, Sheth:2001dp}. The suppression of the number of low-mass haloes occurs for masses below $M_{200} \sim 10^{11} \, h^{-1} \, M_\odot$ and becomes significant at smaller masses. The results obtained with both  functions, Eq.~(\ref{eq:HMFWDM}) (solid red line) and Eq.~(\ref{eq:HMFIDM}) (solid black line) are  displayed for comparison. However, through the rest of the paper we will use Eq.~(\ref{eq:HMFIDM}), since this is a more correct description of the halo mass function in the IDM scenario we consider. Nevertheless, the differences are negligible, even if Eq.~(\ref{eq:HMFWDM}) does not properly account for the smallest resolved structures. Note that for masses below $M_{200} \sim 10^9 \, h^{-1} \, M_\odot$ the contribution of both these fits to the halo mass function to the enhancement factor is completely subdominant. This indicates that, under our assumption about the behaviour of the concentration parameter at small masses (see next subsection), the contribution of the halo mass function to the final result mainly comes from the scales which are resolved in the simulations. Thus, our mass extrapolations actually induce very small uncertainties.

\subsection{Density profile and concentration}

\begin{figure}[t]
	\centering	
	\includegraphics[width=0.5\textwidth]{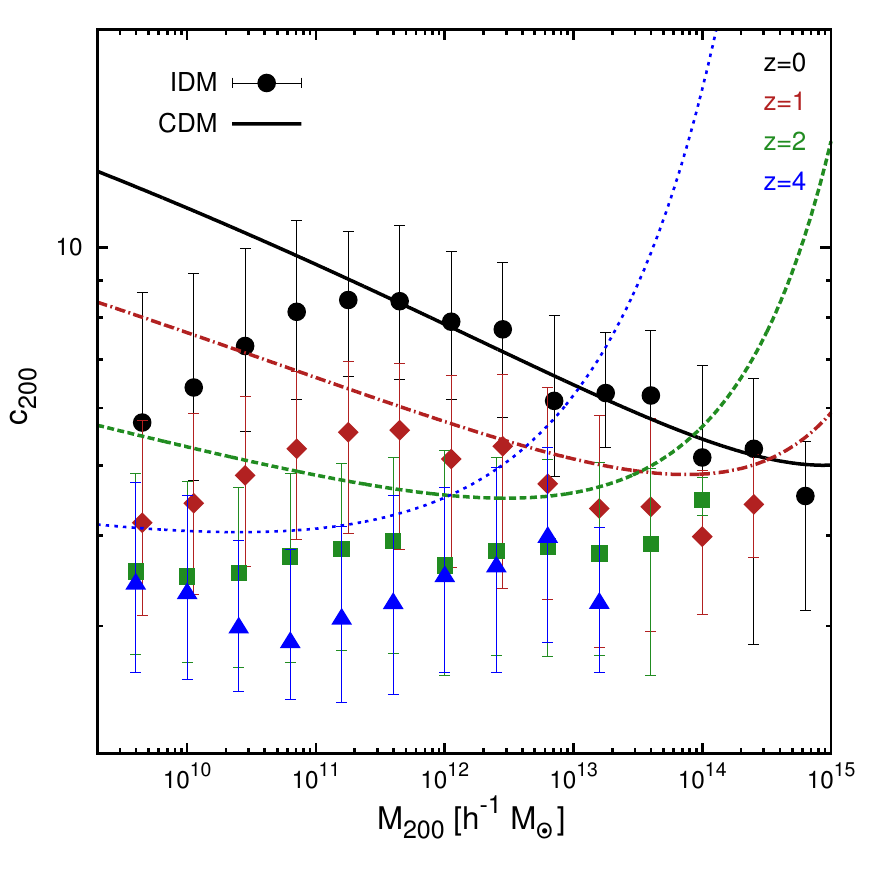}
	\caption{Concentration parameter as a function of halo mass for four different redshifts, $z=0,1,2,4$ (from top to bottom at small masses). The coloured symbols represent the results from the IDM simulations with an elastic DM-$\gamma$ (or similarly for DM-$\nu$) cross section $\sigma_{\rm el} = 2.0 \times 10^{-9} \, (m_{\rm DM}/\rm{GeV}) \times \sigma_{\rm T}$, whereas the lines correspond to the CDM parameterisation in Ref.~\cite{Prada:2011jf}. The error bars mark the 20\% to 80\% interval for the scatter in the halo concentration in each mass bin.}
	\label{fig:concentration} 
\end{figure}

As mentioned above, the density profiles of the haloes in our simulations are in good agreement with an NFW profile, which can be characterised by the concentration parameter. In Fig.~\ref{fig:concentration} we  show the concentration-mass relation for four different redshifts (coloured symbols). As discussed in Ref.~\cite{Schewtschenko:2014fca}, the median value of the concentration parameter for IDM is significantly lower in the mass bins below the half-mode mass compared to CDM (coloured lines), which is explained by the delayed formation of low-mass haloes in these alternative scenarios. This effect is more pronounced for larger cross sections~\cite{Schewtschenko:2014fca}. 

Although not statistically significant, we find an upturn at high masses in the concentration parameter at some of the highest redshifts considered, but this is significantly less pronounced than for CDM\footnote{Although for CDM, unrelaxed haloes in transient stages of their evolution have been claimed as the possible cause of this upturn~\cite{Ludlow:2012wh}.}. However, there are very few haloes at these high masses and consequently they have little impact on the signal.

\subsection{Enhancement factor}

\begin{figure}[t]
	\centering	
	\includegraphics[width=0.5\textwidth]{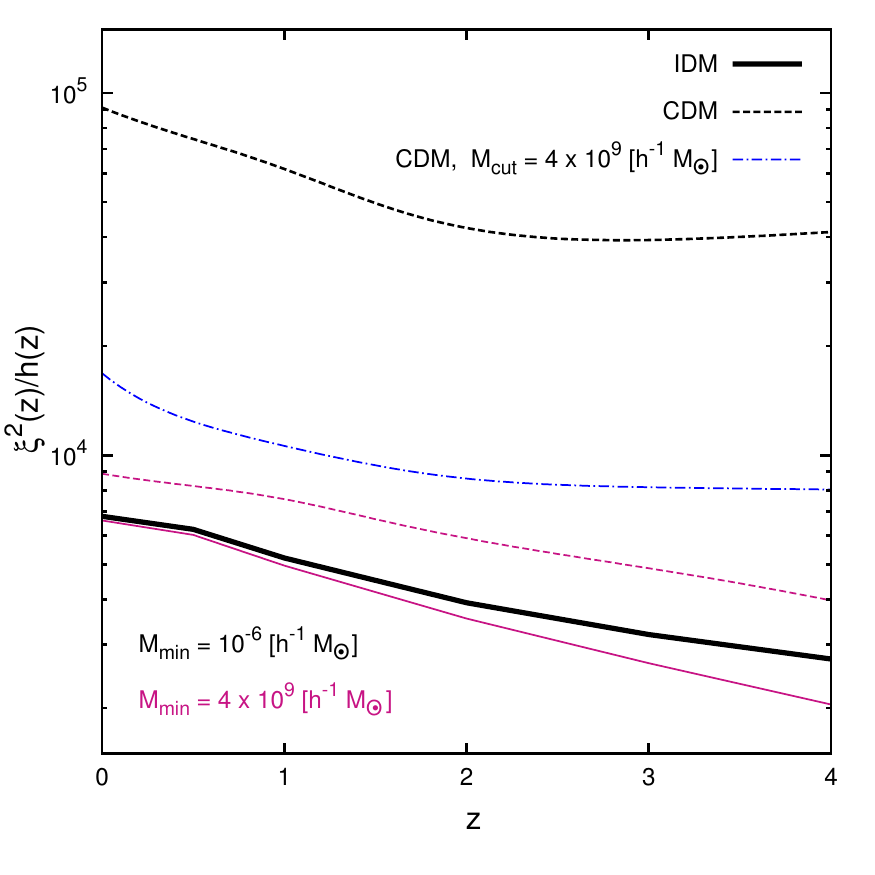}
	\caption{Enhancement factor $\xi^2(z)$, divided by $h(z)$, as a function of redshift for IDM with a  DM-$\gamma$ (or similarly for DM-$\nu$) elastic cross section $\sigma_{\rm el} = 2.0 \times 10^{-9} \, (m_{\rm DM}/\rm{GeV}) \times \sigma_{\rm T}$ and for CDM. Results are shown for $M_{\rm min} = 10^{-6} \, h^{-1} \, M_\odot$ (black lines) and  $M_{\rm min} =  4 \times 10^{9} \, h^{-1} \, M_\odot$ (magenta lines). The solid lines are for IDM and the dashed lines for CDM. For comparison, we also show the enhancement factor for CDM when the concentration parameter is kept constant below $M_{\rm cut} =  4 \times 10^{9} \, h^{-1} \, M_\odot$  (dot-dashed blue line).}
	\label{fig:enhancement} 
\end{figure}

\begin{figure*}
	\centering	
	\includegraphics[width=0.495\textwidth]{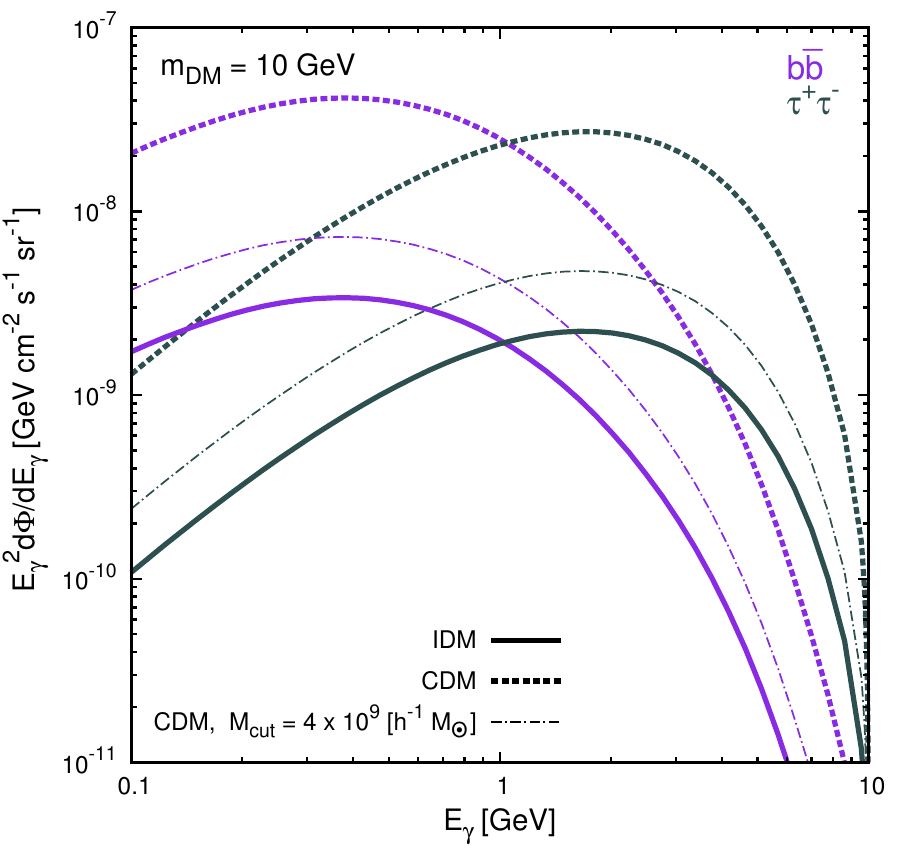}
	\includegraphics[width=0.495\textwidth]{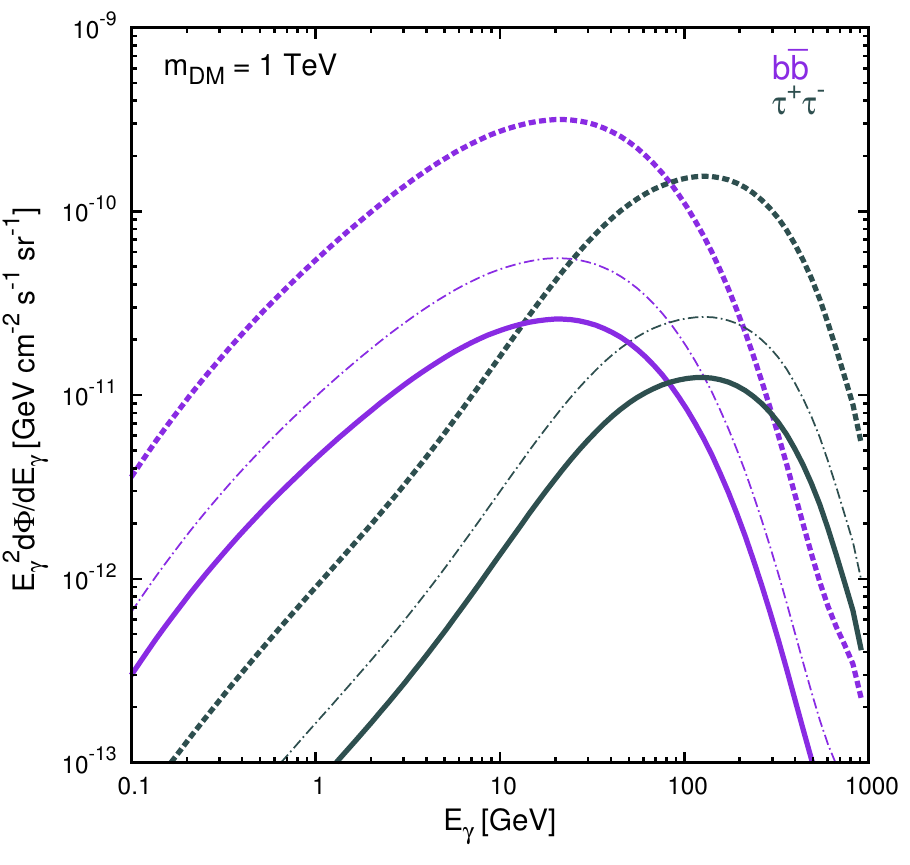}
	\caption{Isotropic extragalactic $\gamma-$ray fluxes from DM annihilations into $b\bar{b}$ (purple lines) and $\tau^+\tau^-$ (black lines) with $\left<\sigma v\right> = 3 \times 10^{-26} \, {\rm cm}^3/{\rm s}$. The left panel shows estimates for a DM mass of 10~GeV while we considered a mass of 1~TeV in the right panel.  The fluxes for IDM are represented by the thick solid lines (assuming an elastic DM-$\gamma$ cross section $\sigma_{\rm el} = 2.0 \times 10^{-9} \, (m_{\rm DM}/\rm{GeV}) \times \sigma_{\rm T}$) and those for CDM by the thick dashed lines. The thin dot-dashed lines represent the CDM case when the concentration is kept constant for masses below $M = 10^{9.6} \, h^{-1} \, M_\odot$. In all cases, we take the minimum halo mass to be $M_{\rm min} = 10^{-6} \, h^{-1} \, M_\odot$.}
	\label{fig:fluxgamma} 
\end{figure*}

As discussed in previous sections, the enhancement in the isotropic extragalactic $\gamma-$ray and neutrino flux from DM annihilations depends on the number density of haloes of all masses at all redshifts and on their internal properties, i.e., their concentration parameter. However, we only have access to these quantities over a limited mass range from the simulations, so we have to rely on extrapolations to lower masses. For the halo mass function we use Eq.~(\ref{eq:HMFIDM}), which was obtained from a fit to the simulation results. On the other hand, in Ref.~\cite{Schneider:2014rda} an upturn in the concentration parameter at low masses was found for WDM scenarios with suppressed small-scale perturbations, which is also hinted from our results at $z=4$ (Fig.~\ref{fig:concentration}). However, in the absence of more conclusive results, we take the concentration parameter to be constant below the minimum halo mass for which we can determine the halo density profile, i.e., $M_{\rm  cut} = 4 \times 10^9 \, h^{-1} \, M_\odot$.

In Fig.~\ref{fig:enhancement} we show the enhancement factor $\xi^2(z)$ (divided by $h(z)$) as a function of redshift for IDM and CDM using different assumptions. We depict the results obtained using the extrapolations of the halo mass function and the concentration down to a minimum mass $M_{\rm min} = 10^{-6} \, h^{-1} \, M_\odot$, for IDM (solid black line) and for CDM (dashed black line). As one can see, the enhancement factor for CDM is an order of magnitude larger than for IDM. We also show the enhancement factor computed using only the halo mass range fully resolved in the simulations, i.e., $M_{\rm min} = 4 \times 10^{9} \, h^{-1} \, M_\odot$ for IDM (solid magenta line) and for CDM (dashed magenta lines). From the comparison of the two black and two magenta lines between themselves, we note that the contribution of low-mass haloes is very important in the case of CDM, while it is subdominant for IDM, which is expected from the suppression of power at small scales. Therefore, the result for the IDM case is quite stable with respect to extrapolations, unless the behaviour of the concentration parameter at small scales turns out to be similar to the CDM case. This will need to be addressed in the future. Finally, we also show the result for CDM by extrapolating the concentration parameter in the same way as done for IDM (dot-dashed blue line). As can be seen, the combination of the abundance of haloes and the growth of the concentration parameter down to small scales implies a much larger enhancement factor for CDM than for IDM scenarios.

\begin{figure*}
	\centering	
	\includegraphics[width=0.495\textwidth]{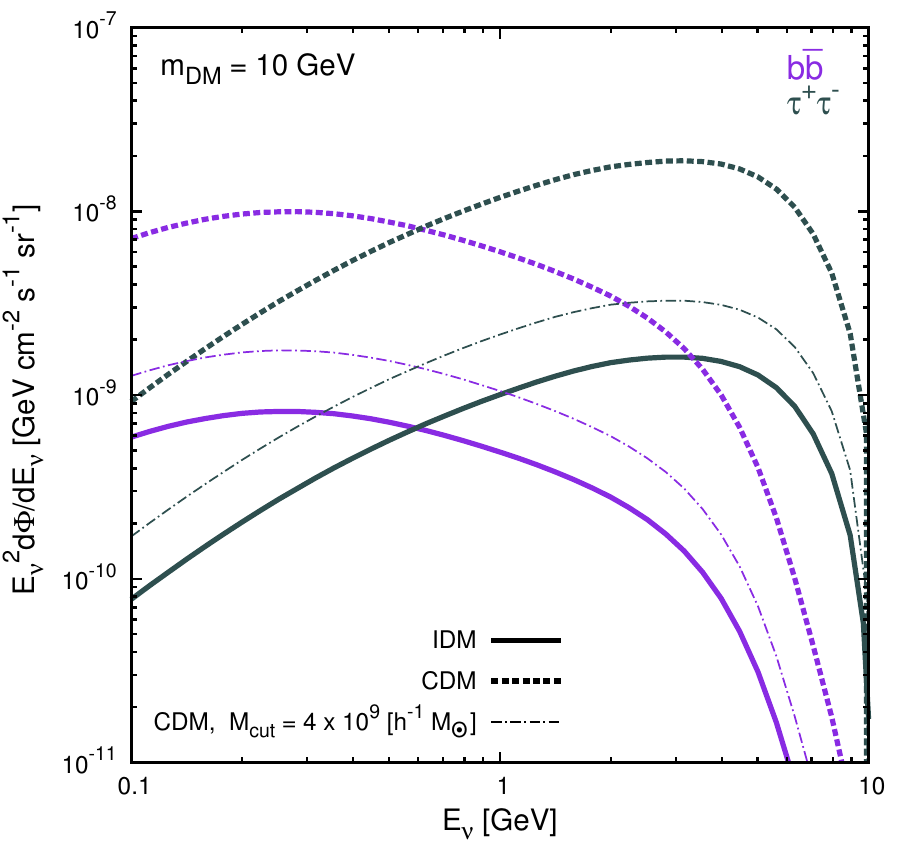}
	\includegraphics[width=0.495\textwidth]{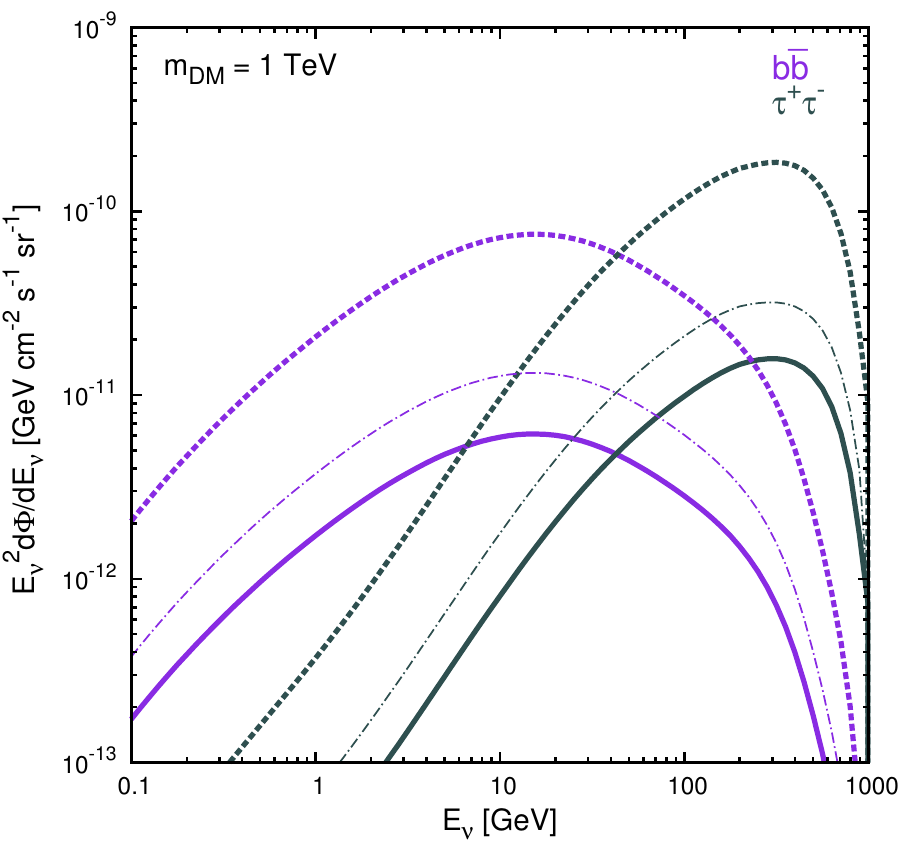}
	\caption{Same as Fig.~\ref{fig:fluxgamma} but for the neutrino fluxes.}
	\label{fig:fluxnu} 
\end{figure*}

\subsection{Extragalactic fluxes}

Now that we have all the necessary ingredients, we can compute the isotropic extragalactic $\gamma-$ray and neutrino fluxes from DM annihilations in IDM scenarios and compare them with the results from the standard CDM case. As done in the rest of the paper, we consider as our default value for the scattering cross section (strictly for DM-$\gamma$ interactions), $\sigma_{\rm el} = 2.0 \times 10^{-9} (m_{\rm DM}/\rm{GeV}) \times \sigma_{T}$.

The results are shown in Figs.~\ref{fig:fluxgamma} and~\ref{fig:fluxnu} for the cases of $\gamma-$ray and neutrino fluxes, respectively. In both figures we assume $\left<\sigma v\right> = 3 \times 10^{-26} \, {\rm cm}^3/{\rm s}$ and consider two DM annihilation channels: annihilations into $b\bar{b}$ (purple lines) and $\tau^+\tau^-$ (black lines), as representative of soft and hard spectra, respectively. In the left (right) panels we consider a DM mass of 10~GeV (1~TeV). We show the expected fluxes for the IDM scenario (thick solid lines) and for the standard CDM case (thick dashed lines). In all cases we notice that in IDM models, the effect of the suppression of the power spectrum at small scales is an overall reduction in the flux with respect to that expected in the standard CDM scenario, which can be of about an order of magnitude for the chosen value of the elastic scattering cross section. This suppression is approximately constant over the entire energy range. Therefore, in case of a signal, the similarities between the energy spectra combined with the flux suppression in IDM scenarios could lead to the misinterpretation of possible evidence for models beyond $\Lambda$CDM as being due to CDM particles annihilating with a smaller cross section than the true one. Let us stress that, whereas the uncertainties associated with the modeling of the cosmological enhancement factor in the standard CDM scenario are expected to be of a factor of a few (see, e.g., Ref.~\cite{Moline:2014xua}), the suppression in IDM scenarios can be much larger.

To determine the impact of the concentration parameter on the flux, we also  estimate the $\gamma-$ray and neutrino fluxes in the CDM scenario by assuming the same extrapolation of the concentration parameter at low masses as in the IDM scenario (thin dot-dashed lines in Figs.~\ref{fig:fluxgamma} and~\ref{fig:fluxnu}).  We observe that, using a constant concentration for masses below $M_{\rm cut} = 4 \times 10^9 \, h^{-1} \, M_\odot$, the CDM fluxes are only slightly greater than the IDM fluxes. Therefore we conclude that the larger values of the $\gamma-$ray and neutrino fluxes in CDM with respect to IDM are not due to the larger number of low-mass haloes, but  rather to the behaviour of the concentration parameter for CDM, which grows with decreasing halo mass unlike the extrapolation we have assumed for IDM.

 \vspace{-1mm}

%%%%%%%%%%%%%%%%%%%%%%%%%%%%%%%%%%%%%%%%%%%%%%%%%%%%%
%%%%%%%%%%%%%%%%%%%%%%%%%%%%%%%%%%%%%%%%%%%%%%%%%%%%%
\section{Conclusions}
\label{sec:conclusions}
%%%%%%%%%%%%%%%%%%%%%%%%%%%%%%%%%%%%%%%%%%%%%%%%%%%%%
%%%%%%%%%%%%%%%%%%%%%%%%%%%%%%%%%%%%%%%%%%%%%%%%%%%%%

We have investigated the isotropic extragalactic signals expected from DM annihilations into $\gamma-$rays and neutrinos, in models where the linear matter power spectrum is suppressed due to DM interactions with radiation (photons or neutrinos). We have described this signal in general terms in Sec.~\ref{sec:xgal} and discussed IDM models in Sec.~\ref{sec:alternative}. The matter power spectrum for these models shares some similarities with that for WDM, but exhibits, in addition, an oscillatory pattern that is similar to baryonic acoustic oscillations~\cite{Boehm:2014vja}. 

The isotropic extragalactic signal depends on the abundance and on the structure of haloes of all masses and up to redshifts of a few. In order to estimate these fluxes in IDM, we use the results of the simulations performed in Ref.~\cite{Schewtschenko:2014fca}, bearing in mind that these simulations have a limited mass resolution below which, we have to rely on extrapolations. Thus, we used Eq.~(\ref{eq:HMFIDM}) as our the halo mass function, the result of a fit to the simulations data, and assumed a constant concentration parameter for halo masses smaller than $M_{\rm cut} = 4 \times 10^{9} \, h^{-1} \, M_\odot$ (see Secs.~\ref{sec:simulations} and~\ref{sec:results}). However, to assess the effect of either a constant concentration parameter or a cut in the power spectrum or both, we also estimated the fluxes in the standard CDM model, with either a cut-off mass scale or a constant concentration below $M_{\rm cut} = 4 \times 10^{9} \, h^{-1} \, M_\odot$.

Our results, presented in Sec.~\ref{sec:results}, show that the spectra of the extragalactic signal for these IDM models have a similar energy dependence in shape to that expected for CDM. However, the expected flux in the IDM scenario can be very strongly reduced with respect to CDM, due to the suppression of small-scale structures and the behaviour of the concentration parameter. Should DM interact significantly with radiation and such an isotropic signal be detected, it is very likely that it could be misinterpreted as being due to DM annihilations in CDM models with a smaller annihilation cross section. However, using the angular power spectrum of the extragalactic signal or cross correlating that signal with possible DM signatures in the Milky Way might actually show inconsistencies and point towards a deviation from CDM.

With this possible degeneracy in mind, detecting such a signal could actually be a new avenue to probe DM interactions with Standard Model particles.

\section*{Acknowledgments}
We thank A. Ibarra for useful comments. AM has been supported by the Funda\c{c}\~ao para a Ci\^encia e a Tecnologia (FCT) of Portugal and thanks the IPPP and IFIC for hospitality. SPR is supported by a Ram\'on y Cajal contract, by the Spanish MINECO under Grants FPA2014-54459-P and SEV-2014-0398, and by the Generalitat Valenciana under Grant PROMETEOII/2014/049. CB and SPR are also partially supported by PITN-GA-2011-289442-INVISIBLES and thank the Aspen Center for Physics where part of this work was done. AM and SPR are also partially supported by the Portuguese FCT through the CFTP-FCT Unit 777 (PEst-OE/FIS/UI0777/2013). We made use of the DiRAC Data Centric system at Durham University, operated by the ICC on behalf of the STFC DiRAC HPC Facility (\texttt{www.dirac.ac.uk}). This equipment was funded by BIS National E-infrastructure capital grant ST/K00042X/1, STFC capital grant ST/H008519/1, STFC DiRAC Operations grant ST/K003267/1 and Durham University. DiRAC is part of the National E-Infrastructure.

\bibliography{biblio}
\bibliographystyle{apsrev4-1}

\end{document}